\documentclass[a4paper]{article}

\usepackage{amsfonts,amsmath,amssymb}
\usepackage{graphicx}
\usepackage{hyperref}
\usepackage[font={small}]{caption}

\def\beq{\begin{equation}}
\def\eeq{\end{equation}}
\def\Rmax{R_{\mathrm{max}}}
\def\Rmin{R_{\mathrm{min}}}
\def\D{\mathrm d}
\def\E{\mathrm e}
\providecommand{\keywords}[1]{\textit{Keywords:} #1}

\title{Highly Deformed Non-uniform Black Strings in Six Dimensions}
\author{Michael Kalisch and Marcus Ansorg \\ \ \\
		\textit{\small{Theoretisch-Physikalisches Institut, Friedrich-Schiller-Universit\" at Jena,}} \\ 
		\textit{\small{Max-Wien-Platz 1, D-07743 Jena, Germany}} \\
		\textit{\small{E-mail: michael.kalisch@uni-jena.de, marcus.ansorg@uni-jena.de}}}
\date{}

\begin{document}

%

\maketitle

\begin{abstract}
We construct numerically static non-uniform black string solutions in six dimensions by using pseudo-spectral methods. An appropriately designed adaptation of the methods in regard of the specific behaviour of the field quantities in the vicinity of our numerical boundaries provides us with extremely accurate results, that allows us to get solutions with an unprecedented deformation of the black string horizon. Consequently, we are able to investigate in detail a critical regime within a suitable parameter diagram. In particular, we observe a clearly pronounced maximum in the mass curve, which is in accordance with the results of Kleihaus, Kunz and Radu from 2006. Interestingly, by looking at extremely distorted black strings, we find two further turning points of the mass, resulting in a spiral curve in the black string's phase diagram.  \\  \ \\
\end{abstract}


\section{Introduction and Summary}

When Gregory and Laflamme (GL) discovered the instability of uniform black strings \cite{Gregory:1993vy} (UBS), a search began for a new branch of solutions that emanates from this instability. Such non-uniform black strings (NBS) were constructed first perturbatively \cite{Gubser:2001ac,Sorkin:2004qq,Wiseman:2002zc} and later numerically \cite{Wiseman:2002zc,Kleihaus:2006ee,Sorkin:2006wp,Figueras:2012xj} by several authors. In Ref.~\cite{Gubser:2001ac} a natural measure of the deformation of the black string horizon, 
$\lambda =\frac{1}{2}\Rmin^{-1}(\Rmax-\Rmin)$, 
was introduced, where $\Rmax$ ($\Rmin$) is the maximum (minimum) of the NBS's areal horizon radius along the compact dimension (for $\lambda =0$ we thus have the UBS). 

An interesting issue in this context is the conjectured phase transition of the NBS leading to a localized black hole \cite{Kol:2002xz} (BH), occurring when the black string's horizon has a critical deformation ($\lambda\to\infty$) and pinches off. So far the numerics gave strong evidence in favour of this conjecture \cite{Wiseman:2002ti,Kudoh:2004hs,Headrick:2009pv}, although the actual transition point was not reached, see Refs.~\cite{Kol:2004ww,Harmark:2005pp}.

The aim of this work is the construction of highly accurate numerical solutions in order to clarify the controversy about the existence of a maximum in the NBS mass curve (cf. Refs.~\cite{Kleihaus:2006ee,Sorkin:2006wp,Figueras:2012xj}), at least in six dimensions. For this purpose a pseudo-spectral scheme was implemented, with sophisticated adaptations to get satisfactory accuracies in the large $\lambda$ regime. The techniques include the use of several appropriate coordinate mappings, the introduction of multiple domains and the split of each metric function into two parts (near infinity). Also, a large number of grid points near the critical point (located on the horizon) was to be taken. Consequently, we were able to reach values of $\lambda\sim 200$.

The described method provided us with results that not only support those of Ref.~\cite{Kleihaus:2006ee} and therefore confirm the maximum in the mass curve. Moreover, they show two further turning points in this curve, i.e.~in sum we found three turning points for the mass, and likewise for the relative tension, the temperature and the entropy. Consequently, we obtain the beginning of a spiral curve in the NBS phase diagram for large $\lambda$. Since our method ceases to converge for values of $\lambda \gtrsim 200$, possible further turning points are to be discussed elsewhere. 

The present results are restricted to six dimensions. Note however that work is in progress regarding five dimensions. The corresponding problem is numerically much more difficult to deal with since the metric functions show logarithmic type behaviour, which becomes a subtle issue if high accuracy is needed. Another interesting question, to be treated in a forthcoming article, regards the occurrence of a spiral curve in the BH phase diagram, which shows a continuous transition to the NBS spiral. If this occurrence can be confirmed, then the phase transition conjecture would have further evidence. Furthermore, we note that a search for unstable modes corresponding to the turning points in the phase diagram would be illuminating.

\section{Metric and Charges}
\label{sec:Metric_and_Charges}

We consider the static NBS metric in $D$ dimensions and with the background $\mathbb R^{D-2,1}\times \mathbb S^1 $ in the form 
\beq
	\D s^2 = -\E ^{2A(r,z)}f(r)\D t^2 + \E ^{2B(r,z)} \left( \frac{\D r^2}{f(r)} + \D z^2 \right) + \E ^{2C(r,z)}\D \Omega ^2_{D-3}. 
	\label{eq1}
\eeq
The three unknown metric functions $A$, $B$ and $C$ depend on the radial coordinate $r\in [r_0,\infty]$ and the coordinate $z\in [0,L]$ varying along $\mathbb S^1$. With 
$f(r)=1-(r_0/r)^{D-4},$
the horizon resides at $r=r_0$. All functions are periodic in $z$ with period $L$. Additionally, the functions $A$, $B$ and $C$ possess reflection symmetry in $z$ with respect to the coordinate value $z=L/2$. 

For $A\equiv B\equiv C\equiv 0$,  Eq.~\eqref{eq1} describes the UBS. Then, the GL instability occurs, if the size of the compact dimension $L$ exceeds a certain value $L_{\mathrm{GL}}$.

The mass $M$ and relative tension $n$ are obtained from the leading coefficients of the asymptotic behaviour of $A$ and $B$ when $r\to\infty$. In contrast, the temperature $T$ and the entropy $S$ can be read off from horizon values (see, for instance, Ref.~\cite{Kleihaus:2006ee}). These four quantities obey Smarr's formula \cite{Harmark:2003dg}
\beq
	TS=\frac{D-3-n}{D-2}M,
	\label{eq2}
\eeq   
as well as the first law of thermodynamics
\beq
	\D M = T \D S + \frac{nM}{L}\D L.
\eeq

In the following we consider $D=6$. Here the critical value of the size of the compact dimension is $L_\mathrm{GL} = 4.95161542007(1) \cdot r_0$ (see also footnote \ref{foot_a}).

\section{Numerical Method}

The basis of our numerical scheme is the expansion of each function in terms of Chebyshev polynomials. Considering the functions' values on Lobatto grid points (which include the boundaries), we can easily calculate spectral derivatives. To solve a system of differential equations, a Newton-Raphson method is applied.

From Einstein's vacuum field equations one finds the system of partial differential equations together with some boundary conditions (see Ref.~\cite{Wiseman:2002zc} for a detailed discussion). A unique solution can be obtained by scaling physical quantities in terms of appropriate powers of $r_0$ and fixing $L$ (we put $L=L_{\mathrm{GL}}$), and by prescribing a value of the function $B$ on the horizon at $z=L/2$.

By an analysis of linear perturbations around the UBS\footnote{\label{foot_a}We set $A=\varepsilon a$, $B=\varepsilon b$, $C=\varepsilon c$ for some small $\varepsilon$ and considered the first order in $\varepsilon$. The analysis of the linear problem also reveals the value of $L_\mathrm{GL}$ (see appendix in Ref.~\cite{Kol:2004pn}) and provides an initial guess for the Newton-Raphson scheme in the non-linear regime. We determined $L_{\mathrm{GL}}$ (in units of $r_0$) up to a precision of $10^{-11}$, cf. section \ref{sec:Metric_and_Charges}.} we find that the following ansatz leads to a rapid decay of the spectral coefficients:
\begin{eqnarray}
	A(r ,z) =& A_0(r) \cdot \left( \frac{r_0}{r} \right) ^2 & + A_1(r ,z) \cdot \cos\left( \tfrac{2\pi}{L}z\right) \cdot \E ^{-\frac{2\pi}{L}r} \cdot \left( \tfrac{r_0}{r} \right) ^{3/2} ,  \nonumber \\
	B(r ,z) =& B_0(r) \cdot \left( \frac{r_0}{r} \right) ^2 & + B_1(r ,z) \cdot \cos\left( \tfrac{2\pi}{L}z\right) \cdot \E ^{-\frac{2\pi}{L}r} ,									           \label{eq3}\\ \nonumber
	C(r ,z) =& C_0(r) \cdot        \frac{r_0}{r}            & + C_1(r ,z) \cdot \cos\left( \tfrac{2\pi}{L}z\right) \cdot \E ^{-\frac{2\pi}{L}r} \cdot  \tfrac{r_0}{r}  . 					   \end{eqnarray}
According to this ansatz, we have to solve for three functions $A_1$, $B_1$ and $C_1$, which depend on the two coordinates $r$ and $z$. Moreover, the functions $A_0$, $B_0$ and $C_0$, depending merely on the radial coordinate $r$,  are to be found. A particular benefit of the decomposition (\ref{eq3}) is the fact that $M$ and $n$ follow from the asymptotics of $A_0$ and $B_0$ alone.

In addition to the aforedescribed split, we introduce new coordinates $\xi$ and $u$ via the transformations 
$r_0/r=\xi ^2(2-\xi )^2$ and $u=\cos(2\pi z/L)$, in which the horizon is located at $\xi =1$ and infinity at $\xi =0$, and where $z=L/2$ corresponds to $u=-1$ and $z=0$ to $u=1$. Note that the term $(r_0/r)^{3/2}$ appearing in (\ref{eq3}) is regular with respect to $\xi$. Also, all $\xi$-derivatives of the occurring functions vanish, when taken at the horizon $\xi =1$.

\begin{figure}[ht]
	\begin{center}
		\includegraphics[scale=0.99]{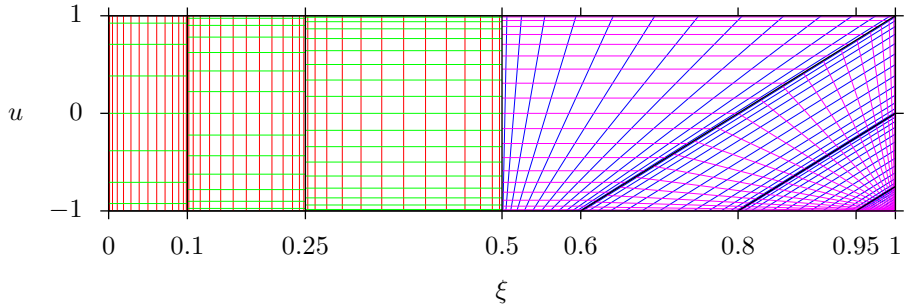}
	\end{center}
	\caption{We divide the integration domain up into seven sub-domains. For $\xi \leq 0.5$ we solve the equations with the ansatz \eqref{eq3}. Here, the three sub-domains guarantee a fast convergence of the spectral coefficients in each of these domains (this subdivision helped to deal with the non-analytic exponential terms appearing in \eqref{eq3}, which occur in higher orders in the non-linear regime). Another benefit is that the resolution in $u$-direction can be adapted in each domain (remember that the $u$-dependency is exponentially suppressed for $\xi\to 0$, while for $\xi$ far from zero high resolution in $u$ is needed). For $\xi \geq 0.5$ we solve for $\alpha=\E^{-2A}$, $\beta=\E^{-2B}$ and $\gamma=\E^{2C}$ (see text for explanation) and introduce further coordinate mappings, that lead to convergence of the coordinate lines in the vicinity of the critical point. An enhancement of the resolution is achieved by a further division into  sub-domains.}
		\label{fig:grid}
\end{figure}

With this ansatz\footnote{Note that we utilized the additional split $C_0(\xi )=C_{00}+\xi ^2 C_{01}(\xi )$ which proved to be useful.} we are able to produce accurate solutions in the NBS branch. However, as in the limit of infinitely pronounced horizon deformations (i.e. $\lambda\to\infty$) the values of the functions diverge at the critical point $(\xi ,u)=(1,-1)$, we introduce the auxiliary functions $\alpha = \exp (-2A)$, $\beta = \exp (-2B)$ and $\gamma = \exp (2C)$. They stay finite within the entire integration domain. Nevertheless, steep gradients appear in vicinity of the critical point, for which reason we introduce appropriate coordinate mappings in order to increase the resolution near this point\footnote{We note that the coordinate $u$ is not suitable in this region for large $\lambda$ ($u$-derivatives become very large at the critical point). We cured this issue by going back to the coordinate $z$, or, to be more precise, to a rescaled version of it, $\tilde u = 1-\frac{4}{L}z$.}, see Fig.~\ref{fig:grid}.  

\section{Results}

\begin{figure}[ht]
	\begin{center}
		\includegraphics[scale=0.52]{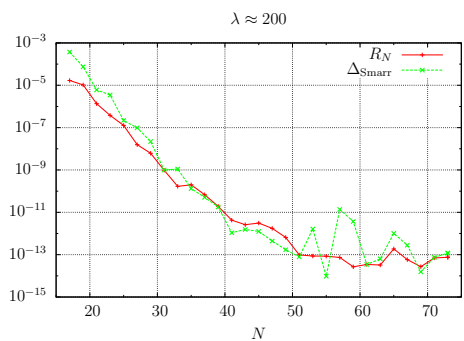}
		\caption{Convergence of the residuum $R_N$ and the deviation from Smarr's formula $\Delta _\mathrm{Smarr}$ for the solution with the highest deformation achieved, $\lambda\approx 200$. We calculate $R_N$ by comparing a solution with high resolution to a solution with smaller resolution $N$ on a fine grid and take the maximal difference.}
		\label{fig:convergence}
	\end{center}
\end{figure}

With the methods described above we are able to construct solutions with the deformation of the black string horizon reaching up to $\lambda\approx 200$. Note that for large $\lambda$, the  accuracy obtained is still of the order $10^{-13}$ (absolute) and $10^{-8}$ (relative). As an additional accuracy check we evaluate Smarr's relation and found agreement of left and right hand sides of (\ref{eq2}) to similar orders (see Fig.~\ref{fig:convergence}). The first law $\D M = T\D S$ (remember $\D L=0$, since $L$ was fixed) is also satisfied to great precision over the whole range of solutions. 

\begin{figure}[ht]
	\begin{center}
	\begin{minipage}{.49\textwidth}
			\includegraphics[width=\textwidth]{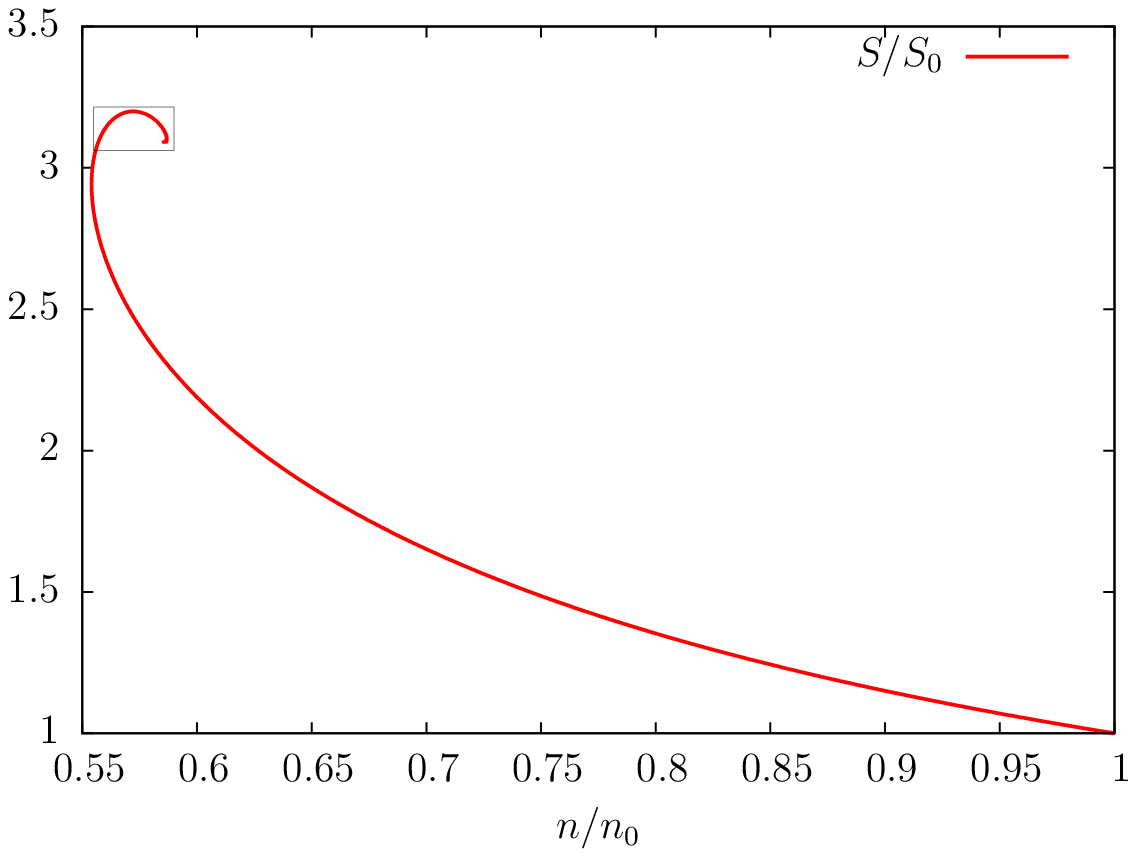}		
	\end{minipage}
	\begin{minipage}{.49\textwidth}
		\includegraphics[width=\textwidth]{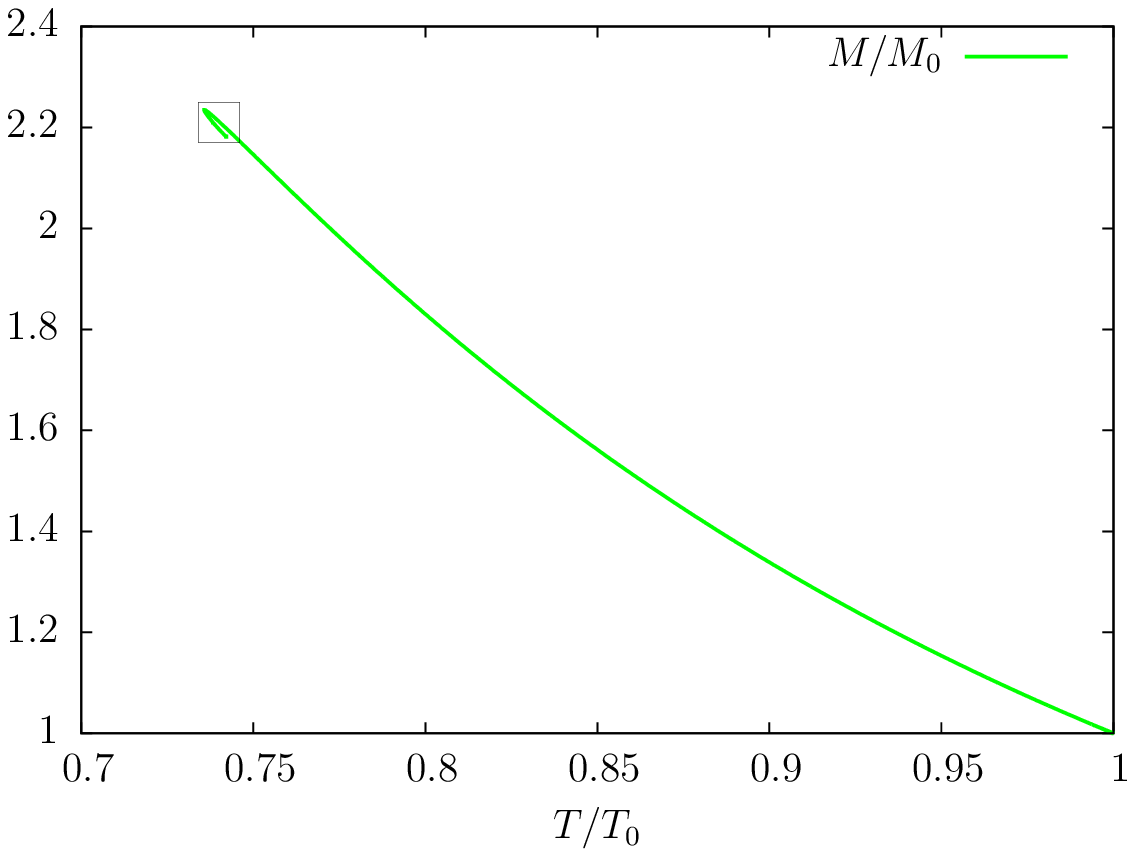}		
	\end{minipage}
	\\	
	\begin{minipage}{.49\textwidth}
			\includegraphics[width=\textwidth]{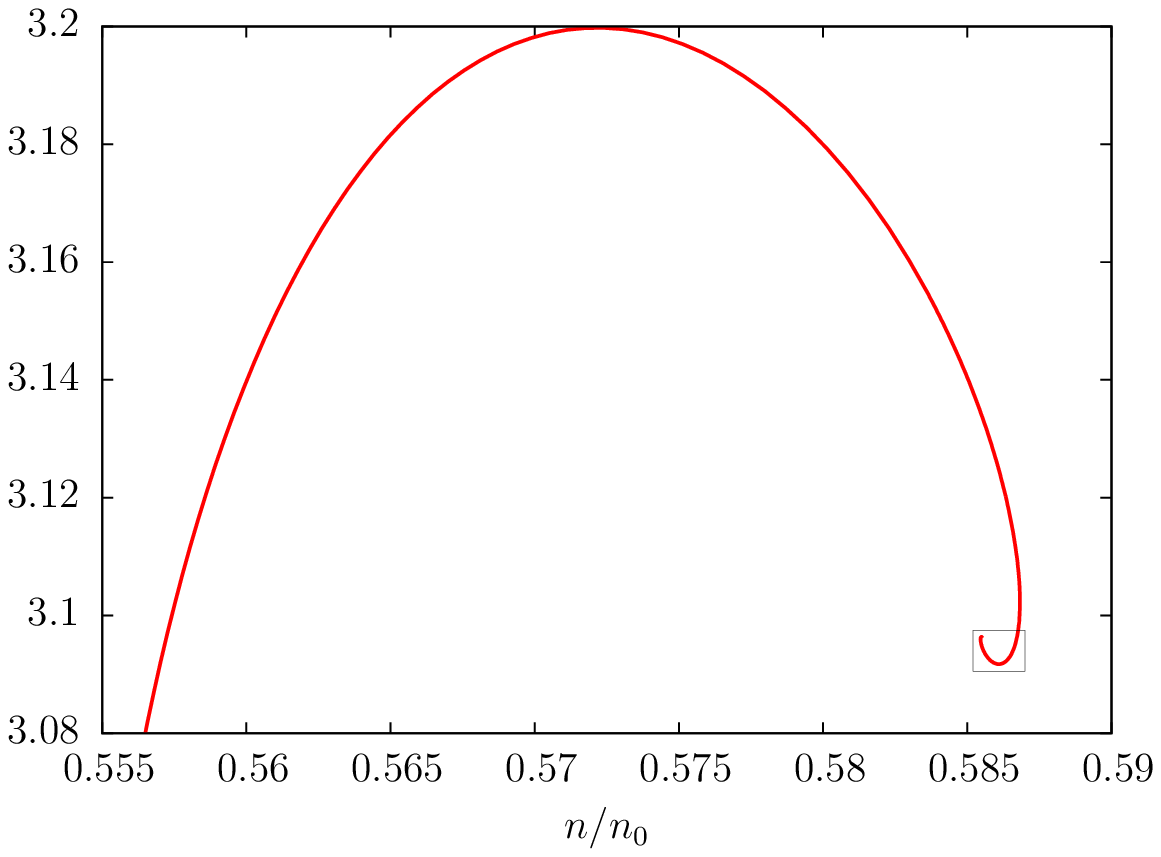}		
	\end{minipage}
	\begin{minipage}{.49\textwidth}
		\includegraphics[width=\textwidth]{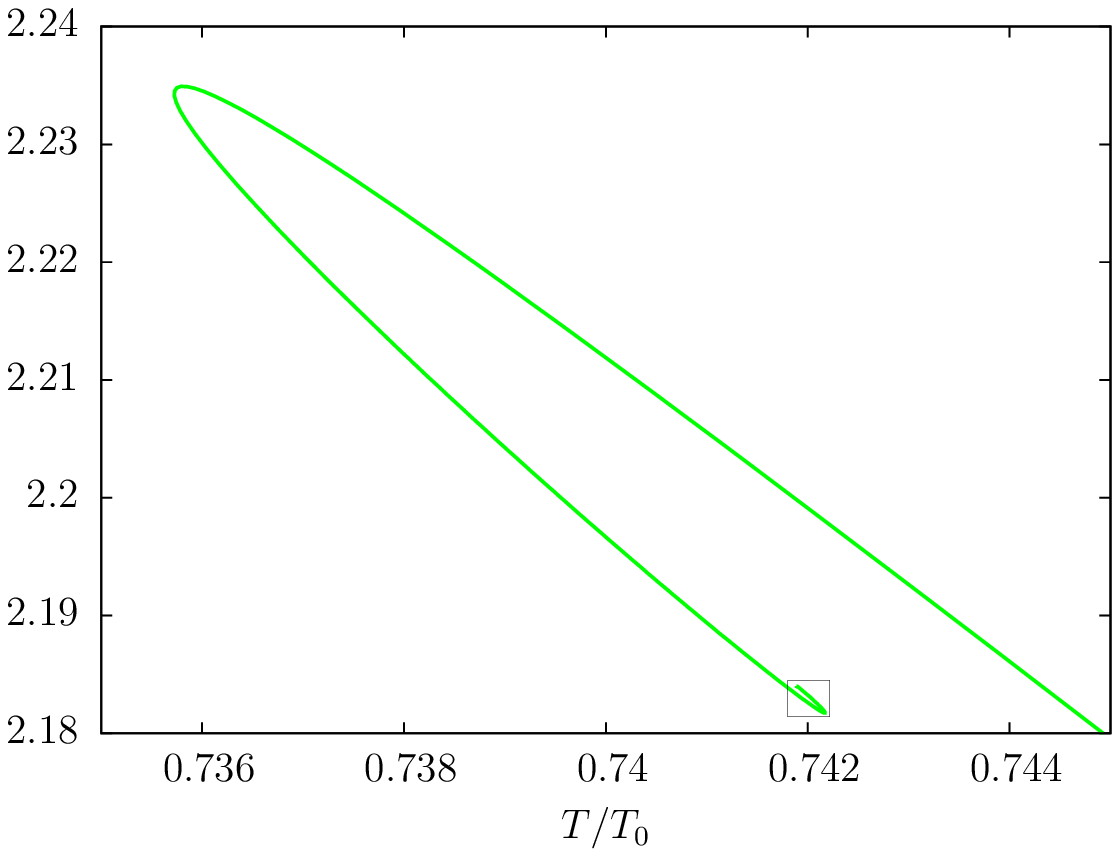}	
	\end{minipage}	
	\\	
	\begin{minipage}{.49\textwidth}
			\includegraphics[width=\textwidth]{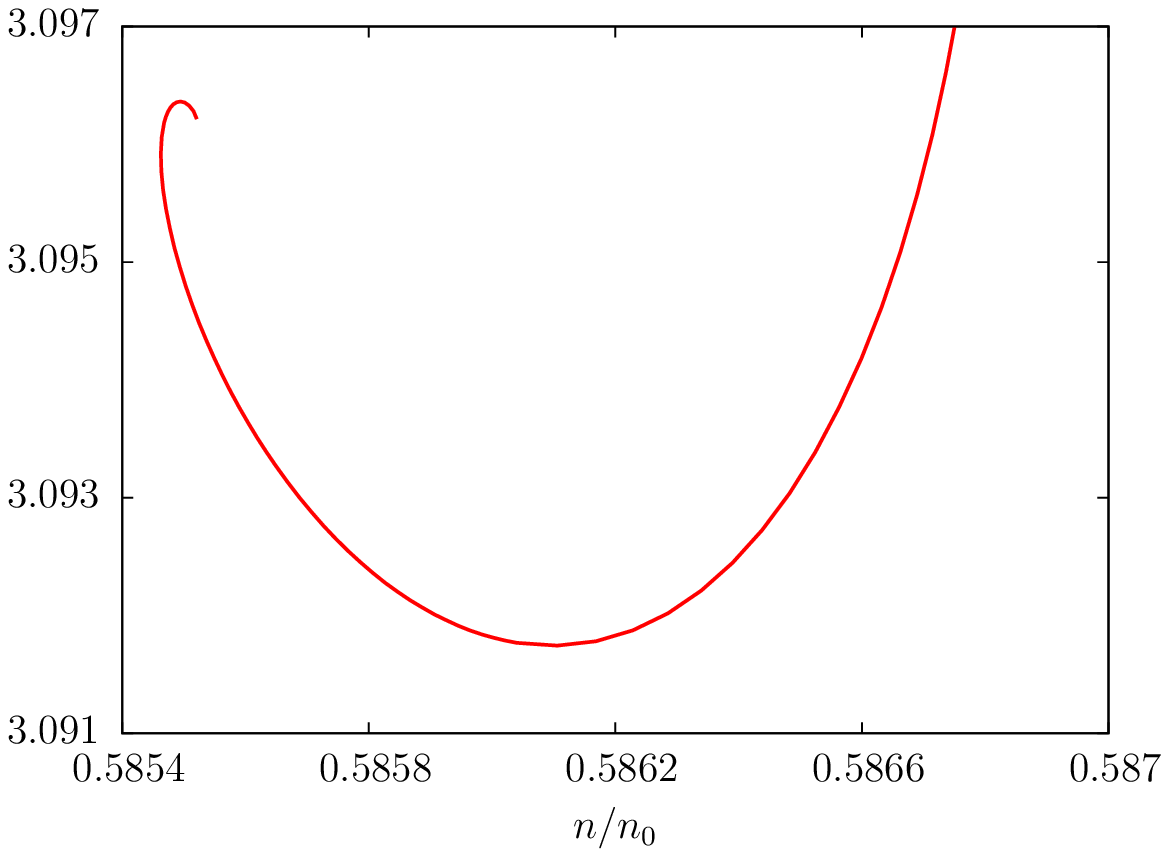}		
	\end{minipage}
	\begin{minipage}{.49\textwidth}
		\includegraphics[width=\textwidth]{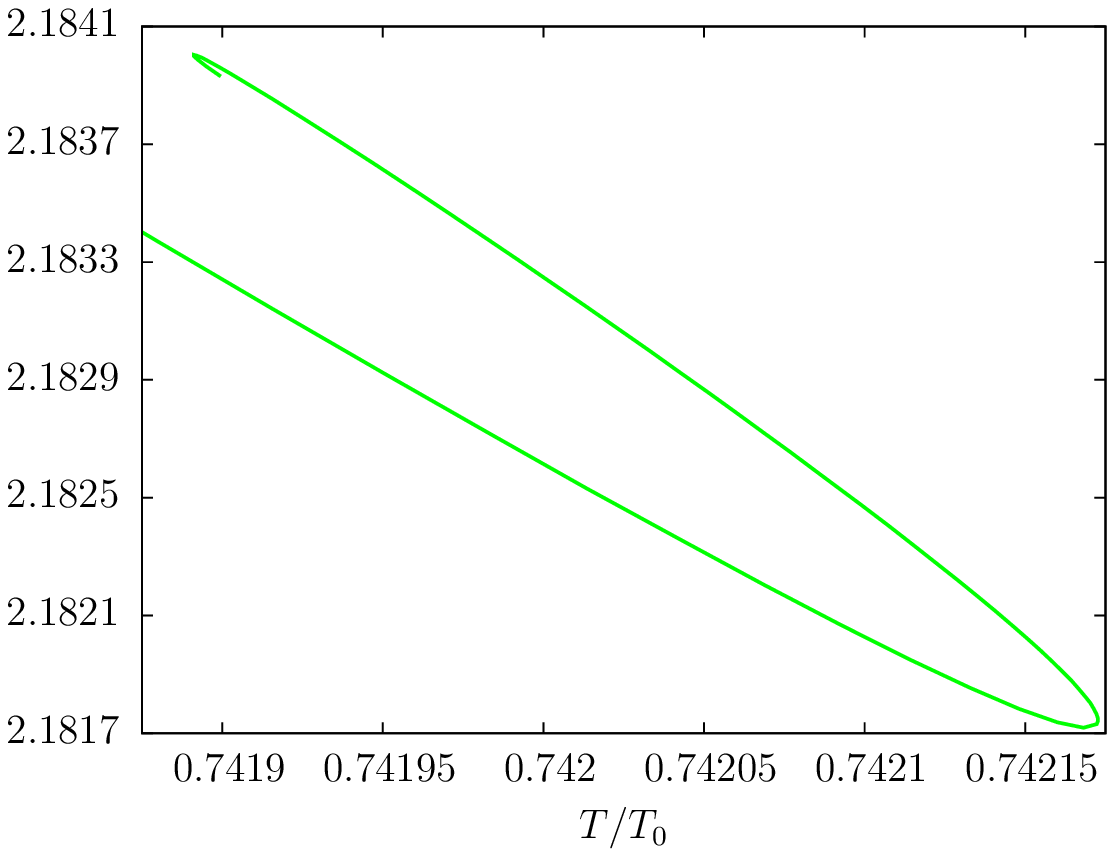}		
	\end{minipage}
	\end{center}
	\caption{Left: Entropy as a function of relative tension. Right: Mass as a function of temperature. The top panels show all of our data. From top to bottom, the panels show more and more magnified portions of the spiral in question. All values are normalized with respect to their corresponding values of the UBS (denoted with lower index $0$).}
	\label{fig:phasediagram}
\end{figure}

The high accuracy achieved permits the detailed consideration of particular black strings' phase diagrams, see Fig.~\ref{fig:phasediagram}. Our results are in very good agreement with those of Ref.~\cite{Kleihaus:2006ee}, which show a maximum in the mass curve (as well as in the entropy curve, whereas relative tension and temperature respectively show a minimum). Going beyond these extremal points by increasing the degree of deformation, another two turning points appear in each of these diagrams. We thus encounter the beginning of a spiral curve in the black strings phase diagrams. Particularly,  in Fig.~\ref{fig:phasediagram} a spiral with about one and a half turns in the $(n,S)$ and $(T,M)$ diagram can be seen (though it becomes tiny). It is tempting to speculate that there are infinitely many turns when going to $\lambda\to\infty$, which would indicate a growing number of unstable modes. 

\newpage
\section*{Acknowledgments}

We thank Burkhard Kleihaus, Jutta Kunz and Eugen Radu for drawing our attention to this problem and for fruitful discussions. Furthermore, we are grateful to Barak Kol for valuable discussions. This work is supported by the Deutsche Forschungsgemeinschaft (DFG) graduate school GRK 1523/2.

\bibliographystyle{unsrt}
\bibliography{NBS}

\end{document}